\documentclass[review]{elsarticle}

\usepackage{lineno,hyperref}
\usepackage{amssymb}
\usepackage{graphics}
\usepackage{graphicx}
\usepackage{epstopdf}
\usepackage{subfigure}

\journal{Journal of \LaTeX\ Templates}

\begin{document}

\begin{frontmatter}

\title{Long-range Correlation and Market Segmentation in Bond Market}


\author[mymainaddress,mysecondaryaddress]{Zhongxing Wang}

\author[mymainaddress]{Yan Yan\corref{mycorrespondingauthor}}
\cortext[mycorrespondingauthor]{Corresponding author}
\ead {yanyan@ucas.ac.cn}

\author[mythirdaddress]{Xiaosong Chen}

\address[mymainaddress]{School of Economics and Management, University of Chinese Academy of Sciences, Beijing, 100080, PR China}
\address[mysecondaryaddress]{Key Laboratory of Big Data Mining and Knowledge Management, Chinese Academy of Sciences,
Beijing, 100191, PR China}
\address[mythirdaddress]{Institute of Theoretical Physics, Chinese Academy of Sciences, Beijing, 100190, PR China}

\begin{abstract}
This paper looks into the analysis of the long-range auto-correlations and cross-correlations in bond market. Based on Detrended Moving Average (DMA) method, empirical results present a clear evidence of long-range persistence that exists in one year scale. The degree of long-range correlation related to maturities  has an upward tendency with a peak in short term. These findings confirm the expectations of fractal market hypothesis (FMH). Furthermore, we have developed a method based on a complex network to study the long-range cross-correlation structure and apply it to our data, and found a clear pattern of market segmentation in  the long run. We also detected the nature of long-range correlation in the sub-period 2007 to 2012 and 2011 to 2016.  The result from our research shows that long-range auto-correlations are decreasing in the recent years while long-range cross-correlations are strengthening.
\end{abstract}

\begin{keyword}
\texttt{long-range correlation}\sep interest rates \sep fractal market hypothesis \sep market segmentation
\end{keyword}

\end{frontmatter}


\section{Introduction}

As a measure of cost of capital and profit in modern financial market, interest rate is top priority for any country. Interest rates have become the core content of the economic research not only for its nature of value measurement in bond market, but also for its importance in monetary policy transmission. Changes in interest rates generally reflect the tightness of the monetary policy of a country. In this way, interest rates significantly influence money supply, lending, stock market, and real economy in the end \cite{fabozzi2007fixed}.\\

Earlier on, interest rates (prices of bonds) and prices of other financial properties are believed to follow random walk \cite{smith1976option, malkiel1970efficient}. Thus price changes are assumed to obey Gaussian distributions. However, recent researches prove that price changes follow a complex distribution with a more obvious peak and fatter tails than Gaussian's \cite{peters1994fractal, mantegna1999introduction, di2001empirical}. It states in some economists' and physicists' research that interest rates fluctuation shows some complex properties, such as long range correlation or memory \cite{Tabak2005The,Cajueiro2007Time, McCarthy2004}, fractals/multifractals \cite{cajueiro2007long, Wang2016Multifractal}, and so on.\\
Long-range correlations appear in several kinds of equities in literature \cite{Jiang2007Non, He2010Are, Cajueiro2005Testing}, but only a few of them focus on bond market. Backus and Zin (1993) \cite{backus1993long}seem to be the first to consider the existence of long-range memories in interest rates. They use a fractional difference model to study the features of yields on US government bonds with modern asset pricing theory and find the evidence of long memory in the 3-month zero-coupon rate. Cajueiro and Tabak (2006) \cite{cajueiro2007long}test for long-range dependence in the term structure of London Interbank offered rates (LIBOR), which are considered to be the benchmark interest rates in European countries. They found significant evidence which shows that interest rates have a strong degree of long-range dependence and furthermore a multifractal nature. And they also suggest that the pricing of interest rates derivatives and fixed income portfolio management should take long memories into account.\\

Overall, only a small number of papers deal with long memory on interest rates.
The main focus in these papers lies on one interest rate or interest rates of one sort, and lack of large sample analysis, which can be considered insufficient in this context. Furthermore, the existing literature mainly concentrates on long-range auto-dependences and neglects the cross-correlations between interest rates. As bond market is a complex system including correlations between various interest rates, shocks in one rate may have a significant influence on others in a short or long time. Market segmentation theory, an important economics thought, suggests that interest rates are divided into several groups according to maturities and lack of substitution between each other. Some literature has found evidence of market segmentation in short run \cite{kidwell1983market, simon1991segmentation, hendershott1978impact, wang2016time}, but the analysis in long-range correlations is still absence.\\
In our work, long-range correlations of interest rates, including auto-correlations and cross-correlations, are analyzed in a relatively large sample, which contains nearly all the important rates in China bond market. Instead of simply providing empirical evidence, we will investigate the underlying mechanisms of economics by statistics and economic analysis. Furthermore, we will develop a long-range memory network to show cross-correlations between interest rates and a clear pattern of market segmentation.  Findings from this research provide evidence to the market segmentation theory and fractal market hypothesis.\\
 This paper is organized as follows: the first part is an introduction of background of our research. The methodology to test long-range correlations will be briefly introduced in part two. Then in section 3 there will be a description of the databases we used in this study followed by discussions about auto-correlations, cross-correlations between interest rates and time line analysis. Finally, we end with a conclusion in section 4.\\

\section{Methodology}

Long-range correlations in time series may be detected in several ways. Two known methods are Rescaled Range Analysis (R/S) \cite{hurst1951long} and Detrended Fluctuation Analysis (DFA) \cite{peng1994mosaic, kantelhardt2001detecting}. However, R/S is seriously biased when the data also exists short range dependence \cite{Willinger1999Stock} and DFA is based on discontinuous polynomial fitting, oscillations in the fluctuation function and significant errors in crossover locations can be introduced \cite{Alvarez2005Detrending}. To overcome the shortcomings of the DFA method, a moving average (MA) process is introduced to replace the polynomial fitting detrending method in DFA by Alessio et al.(2002)\cite{Alessio2002Second}.  For this research paper, we will therefore use Hurst exponent provided by Detrended Moving Average method (DMA) to measure long-range correlation for the priorities of DMA compared to DFA in several numerical experiments \cite{gu2010detrending}.\\

The Detrended Moving Average method (DMA) is similar with Detrended Fluctuation Analysis (DFA) except for the trends fitting function. These two are initially proposed for one time series and can be extended to two time series.
Let us suppose $\left\{R(i)|i=1,2,\dots,N\right\}$ to be the time series of interest rates, where N is the length of the series. The fluctuation is defined as\\
 \begin{equation}
  X(i)=\left|R(i)-R(i-1)\right|
 \end{equation}
The \emph{profile} is given by\\
\begin{equation}
  Y(t)=\sum_{i=1}^t \left(X(i)-\langle X\rangle\right),
  \end{equation}

Divide the profile (length of series is \emph{N}) \emph{Y(t)} into  $N_s=\lfloor N/s\rfloor$ non-overlapping segments of equal length \emph{s}. Since the series length \emph{N} may not be a multiple of the time scale \emph{s}, a proper way is to repeat the method from the opposite end. Thereby, $2N_s$ segments are obtained altogether. And then the local trends $p_v (i)$ are calculated by polynomial fit (DFA) or moving average method (DMA) in this paper. In every segment \emph{v}, we use the original data minus the local trend and get the detrended time series.
\begin{equation}
Y_s(i)=Y(i)-p_v(i)
\end{equation}
Then the variance is given by
\begin{equation}
F_{DFA}^{2}(v)=\langle Y_s^2(i)\rangle=\frac{1}{s} \sum_{i=1}^s\{ Y_s^2[(v-1)s+i]\},
\end{equation}
The detrended variance is calculated as follows
\begin{equation}
F(s)=\frac{1}{2N_s}[\sum_{v=1}^{2N_s}F_{DFA}^2(v)]^{\frac{1}{2}}
\end{equation}
Then the following scaling relation can be expected
\begin{equation}
F(s)\propto s^H
\end{equation}
Through the least-squares fit, the slope of $\mathit{lg}F(q,s)$ and $\mathit{lg} s$ is the Hurst exponent \emph{H}, which is a measurement of long-range memories. In particular, $0<H<0.5$ represents  negative (antipersistence) correlations and $0.5<H<1 $ corresponds to  positive (persistence) correlations, while $H=0.5$ suggests a fully uncorrelated signal\cite{kantelhardt2001detecting}.\\
If there are two time series to be taken into consideration, the DFA or DMA method expands to Detrended Cross-Correlation Analysis (DCCA) method\cite{podobnik2008detrended}, and the variance function is given by
\begin{equation}
    F_{DCCA}^{2}(v)=\langle Y_s(i)\cdot Y'_s(i)\rangle=\frac{1}{s} \sum_{i=1}^s\{ Y_s[(v-1)s+i]\times Y'_s[(v-1)s+i]\},
\end{equation}
where $Y'_s(i)$ is the detrended time series of the other interest rates and the computing process is the same with $Y_s(i)$.\\
To measure the statistically significance of cross-correlation, Zebende (2011) and Podonik et al. (2011)\cite{Podobnik2011Statistical} proposed a detrended cross-correlation coefficient, which is defined as
\begin{equation}
    \rho_{DCCA}(s)=\frac{F_{DCCA}^2(s)}{F_{DFA}(s)F'_{DFA}(s)}
\end{equation}
Here $\rho_{DCCA}(s)$ is a dimensionless coefficient ranging from -1 to +1, similar to standard Pearson correlation coefficient.\\

\section{Data and Empirical Analysis}
Our sample is composed by more than 100 major interest rates published daily by China Central Depository and Clearing Company (CCDC), which includes nearly all types of bonds traded in the interbank market. Built in 1997, the interbank market is the main bond-trading market in China. Over 97\% transaction of bonds are happening in interbank market. What's more, Our sample contains not only short-term monetary rates, but also long-term rates, such as corporate bonds rates and policy financial bonds rates, etc. (see Table.1. 1d means overnight in our research).\\
Since the study is on the long-range property in time scale, the timeline is supposed to last as long as possible. The study period is from January 4, 2007 to January 11, 2016.\\

\begin{table*}
\centering
\caption{\label{tab:table3}The sample data and its abbreviations of our reasearch}

\begin{tabular}{|p{6cm}|c|c|}
Types&Abbrev.&Maturities\\\hline
Pledged Repo Rate&R&1d-14d\\
Interbank Offered Rate&IBO&1d-14d\\
SHIBOR&SHIBOR&1d-1y\\
Central Bank Bill&CBB&1d-3y\\
AAA-Short-term Note&SN\_AAA&1d-5y\\
AAA-Subprime Rank Commercial Bank Financial Bond&SuCBFB\_AAA&1d-20y\\
Treasury Bond&TB&1d-30y\\
Policy Financial Bond&PFB&1d-30y\\
AAA-Corporate Bond&CB\_AAA&1d-30y\\

\end{tabular}
\end{table*}

\subsection{Long-range Auto-correlation}
As it is presented in the previous part, when time series have a long range self-correlation, $\mathit{lg}F(s)$ and $\mathit{lg}s$ will then have a linear relationship (see Equation(6)). Fig 1. shows the result of power law correlation of our interest rates sample. The linear relationships imply the dynamics insinuated in interest rates does not obey random walk but follows a pattern of power-law relationships in auto-correlation of all interest rates. Most rates are in  good linear relationships  on the condition that the window size $s<102.4\approx251$, which exactly is the amount of trading days in one year. The scale of power-law relationships, or long-range correlations in interest rates are not indefinite but within one year. When the window size is beyond one year, the slopes of most rates are reduced to 0.5, which implies a random walk (but for some short-term rates, the detected slopes are more than 1). So the window size is restricted to one year---250 days when we calculate Hurst exponents using DMA method in the following study.\\

    \begin{figure}
    \centering
      \includegraphics[width=1\textwidth]{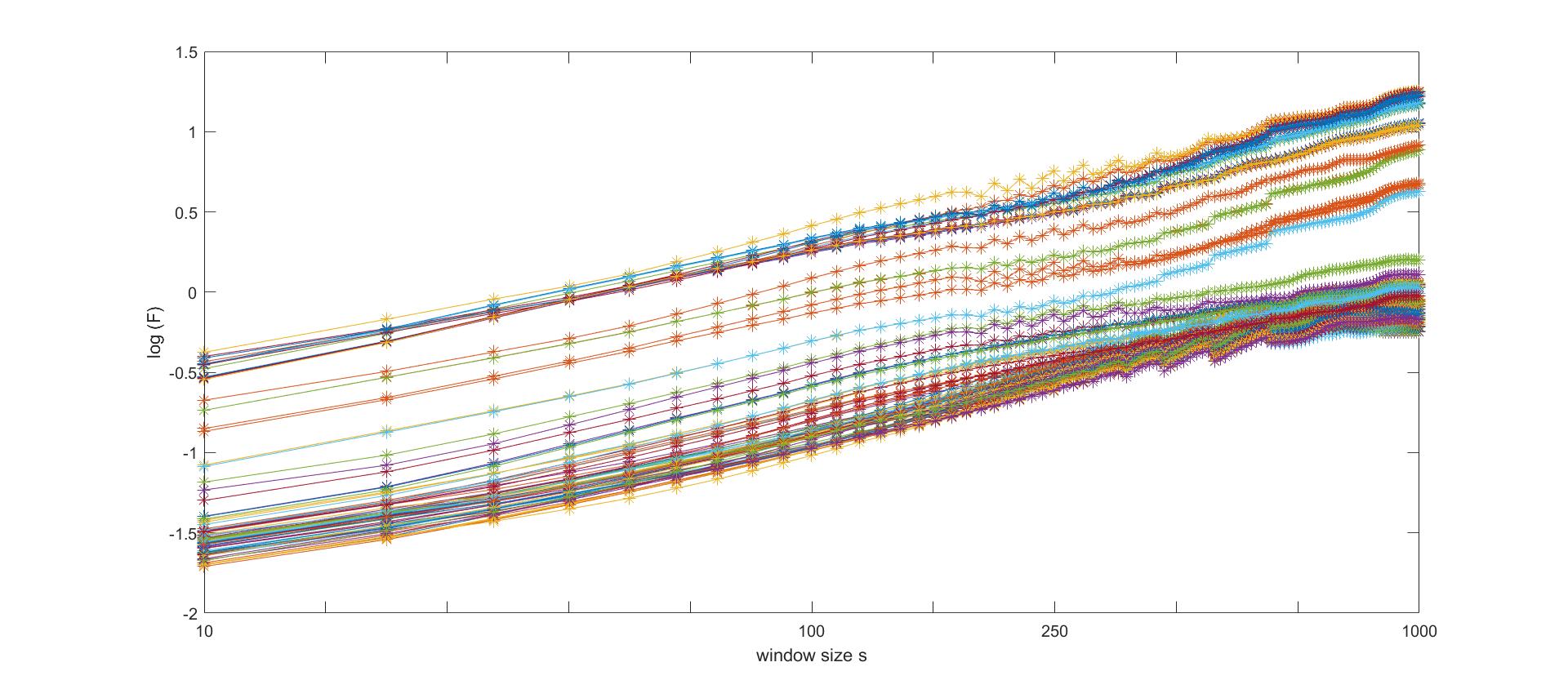}

      \caption{\footnotesize{ The linear relation between fluctuation function and window size. Each line in the figure represents a power-law relationship between fluctuation function and window size, which implies a long-range correlation existence in the interest rate. We can see clearly that most rates are in good linear relationships when window size is smaller than 250 days, which is exactly the amount of trading days in one year.}}

    \end{figure}

Fig.2 shows the distribution of Hurst exponents in our study sample. The Hurst exponents range from 0.63 to 0.95, larger than 0.5, with a peak value around 0.83, which means that the long-range correlation is positive. A sudden shock in interest rates may have a persistent influence on itself in the future. The fluctuation pattern of interest rates is not completely random; a large price wave is more possible to be followed by another large wave rather than a small one.\\
\begin{figure}
    \centering
      \includegraphics[width=1\textwidth]{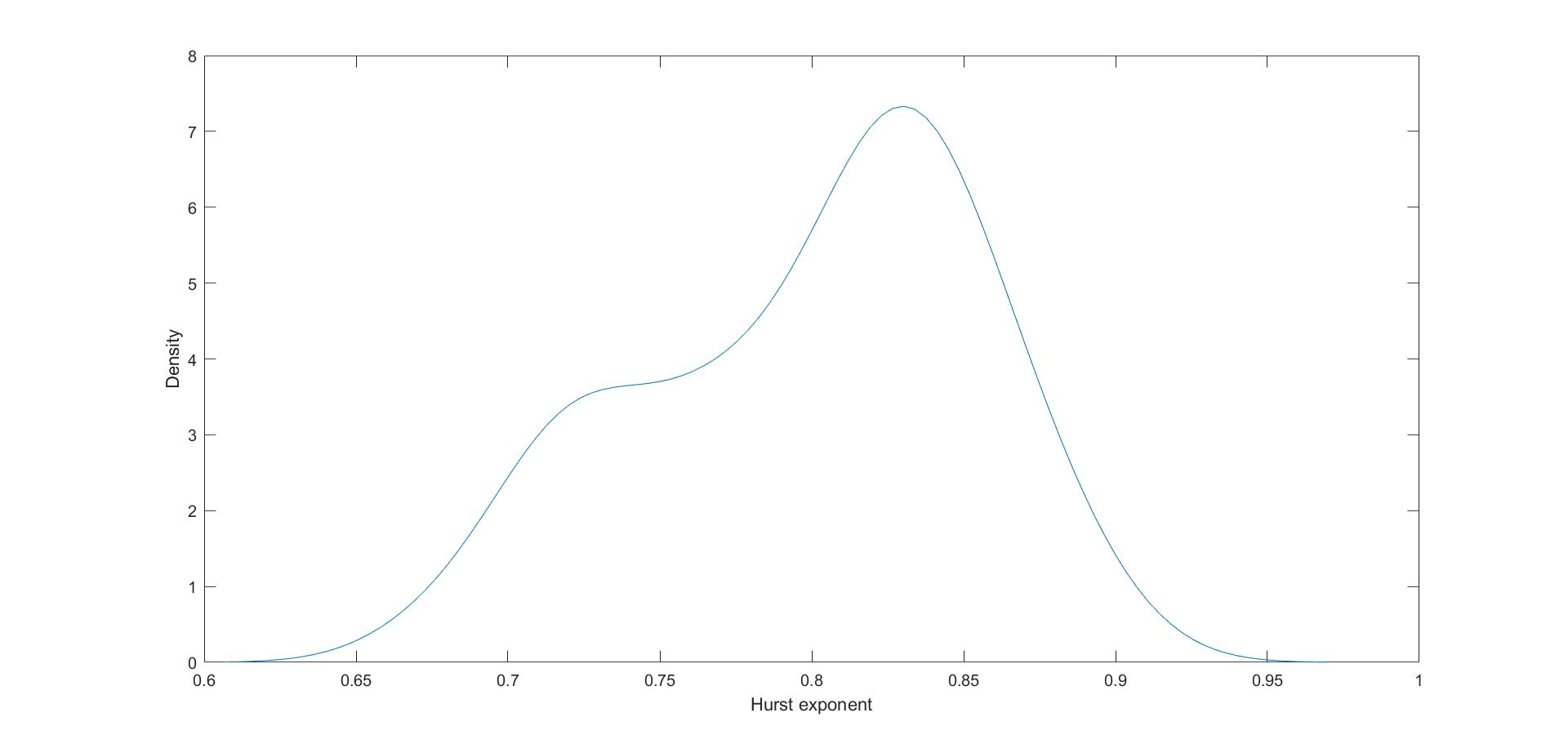}

      \caption{\footnotesize{  The distribution of Hurst exponents. The peak is near 0.83, and most of the exponents are range from 0.63 to 0.95, larger than 0.5, which means a long-range persistence.}}

    \end{figure}

To have a better understanding inside the economic meanings, Hurst exponents can be considered according to the bonds' maturities.\\
The way Hurst exponents of Treasury Bonds (TB), Policy Financial Bonds (PFB), Corporate Bonds (CB\_AAA) change over maturities is shown in Fig.3. Conceptually, Hurst exponents of these three kinds of bonds tend to be larger as maturities increase. Still we can see a vague peak in the maturity of 3 or 6 months.\\
\begin{figure}
    \centering
      \includegraphics[width=1\textwidth]{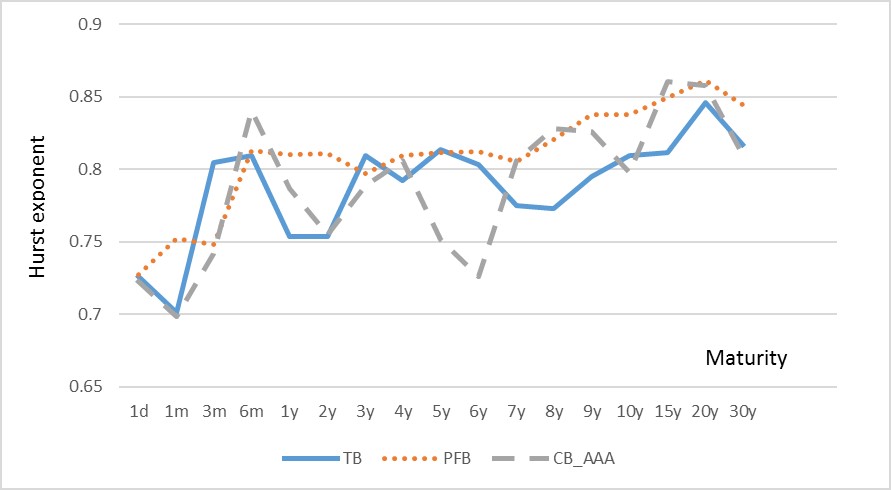}

      \caption{\footnotesize{  Hurst exponents of Treasury Bonds (TB), Policy Financial Bonds (PFB), Corporate Bonds (CB\_AAA) change over maturities. Conceptually, Hurst exponents of the three kinds of bonds have a term structure of an upward tendency, together with one vague peak near 3 months to 6 months term.}}

    \end{figure}

The peak in short-term maturity indicates Hurst exponents of 3/6 months are larger than contiguous terms, is in line with empirical results in US, Europe, Australia and Britain \cite{Cajueiro2007Time}. Larger Hurst exponent means a stronger long-range auto-correlation in time series. When we consider predicting the value of interest rates,  higher exponents signifies a more predictable time series. The reason for this phenomenon is that Central banks in most countries is referring to short-term targets, inflation targets for example, and smooth the target interest rate path to achieve.  In this way, short-term interest rates (which are set up by Central Banks) are more predictable than longer term rates \cite{mishkin2007economics}.\\
However, the upward tendency of Hurst exponents along with the term structure differs from previous work that has been focusing on other countries \cite{Cajueiro2007Time}. Established in 1997, the development of interbank bond market of China has history of only 20 years or so, has not reached its matured stage yet. Bond trading in China is therefore not as active as developed countries, especially in longer terms. Investors in bond market do not like to hold bonds with too long maturities for the fear of the unknown risk in the future. Thus bonds trading with too long maturities are inactive and lack of liquidity. According to fractal market hypothesis (FMH) proposed by Edgar Peters \cite{peters1994fractal}, liquidity is the precondition of market stability, causing the financial market to be turbulent when there is a lack of liquidity. A price shock happening in the market might be the sell-point for some market participants and since the market lacks participants who take this shock as buy-point, the influence will last for a long time, causing a stronger long-range memories of long-term bonds.\\

\subsection{Long-range Cross-correlation Network}
Bond market is believed to be a complex system composed of interest rates that mutually interact in a complex fashion that the current value of each interest rate depends on not only the past value of itself but also the value of other interest rates. Using conditional Granger causality test method, Wang et al.(2016)\cite{wang2016time} constructs a directed causal correlation network to show the cross-correlation structure between interest rates, which is segmented into different groups according to bonds' maturities, confirming the existence of market segmentation in China bond market. However, restricted to the algorithm, Granger causality test can only describe correlations in a rather short run.\\
As mentioned in previous part of the paper, interest rates have a significant long-range auto-dependence. A shock to interest rate will have a long-range persistent influence on itself. We can also expect the influence contagion from one interest rate to others as well as persistency for the long run. Via the detrended cross-correlation coefficient $\rho_{DCCA} (s)$ as mentioned before, a measurement of degree of cross-correlations is given between interest rates.\\

We selected several key interest rates in short, medium, and long term interest rates groups separately---R1d, R7d in short-term group, CBB2Y, TB2Y in mid-term group, and TB7Y, TB10Y in long-term group. According to the work of \cite{wang2016time}, these interest rates are believed to be representative  since they are in the core position of causal network. Long-range correlation coefficients $\rho_{DCCA} (s)$ of the key rates are calculated for different window sizes s ranging from 5 to 500. The result is shown in Fig.4. It is clearly that correlation coefficients of interest rates in the same term group (the top three solid lines) are larger than correlation coefficients between different groups (the bottom two dashed lines) in any window sizes. Long-range correlation between interest rates of same or similar maturities is stronger than correlation between rates far apart in maturities. Influence structure seems to be segmented according to different maturities. Secondly, there appear to be two turning points on each line in Fig.4. One is near window size $s=22$, which is the amount of trading days in one month. The other one is near $s=250$, year scale.\\
\begin{figure}
    \centering
      \includegraphics[width=1\textwidth]{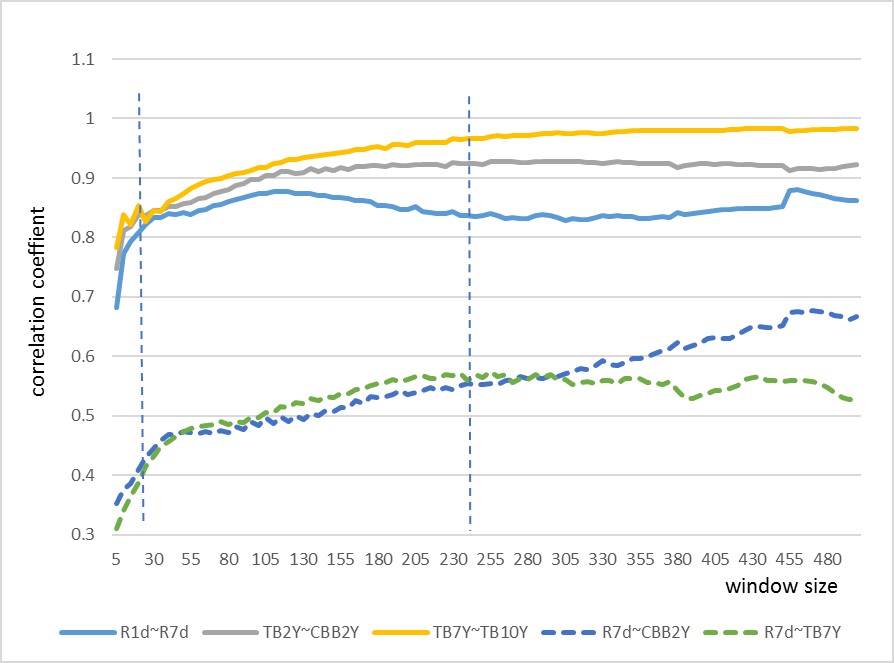}

      \caption{\footnotesize{ Long-range correlation coefficients $\rho_{DCCA} (s)$ of the key rates are calculated for different window size s range from 5 to 500. Correlation coefficients of interest rates in the same term group (the top three solid lines) are larger than correlation coefficients between different groups (the bottom two dashed lines) in any window size. What's more, there appear to be two turning points lie on each line. One is near window size $s=22$ (month scale), the other one is near $s=250$ (year scale).}}

    \end{figure}

In order to show our results intuitively, a network method is applied to represent the correlation structure. Fig.5 shows the long-range correlation network with window size $s=50, 150, and 250$. Nodes in the network represent the interest rates and the edges relates to the long-range correlation described by the value of $\rho_{DCCA} (s)$. To be more concise, we remove the edges whose coefficient absolute values are less than 0.8. A modularity algorithm \cite{blondel2008fast, lambiotte2008laplacian} is used to detect the modules of the networks. It is obvious that the long-range correlation structure segmented into three groups according to the maturity. The top left (represented in red) consists of bonds with maturities less than 6 months. Top right part (represented in blue) of network contains interest rates longer than 3 years. And the rest of network (represented in green) is made of rates longer than half a year but less than 3 years.\\
\begin{figure*}
\centering
\subfigure[$s=50$]{\includegraphics[scale=0.175]{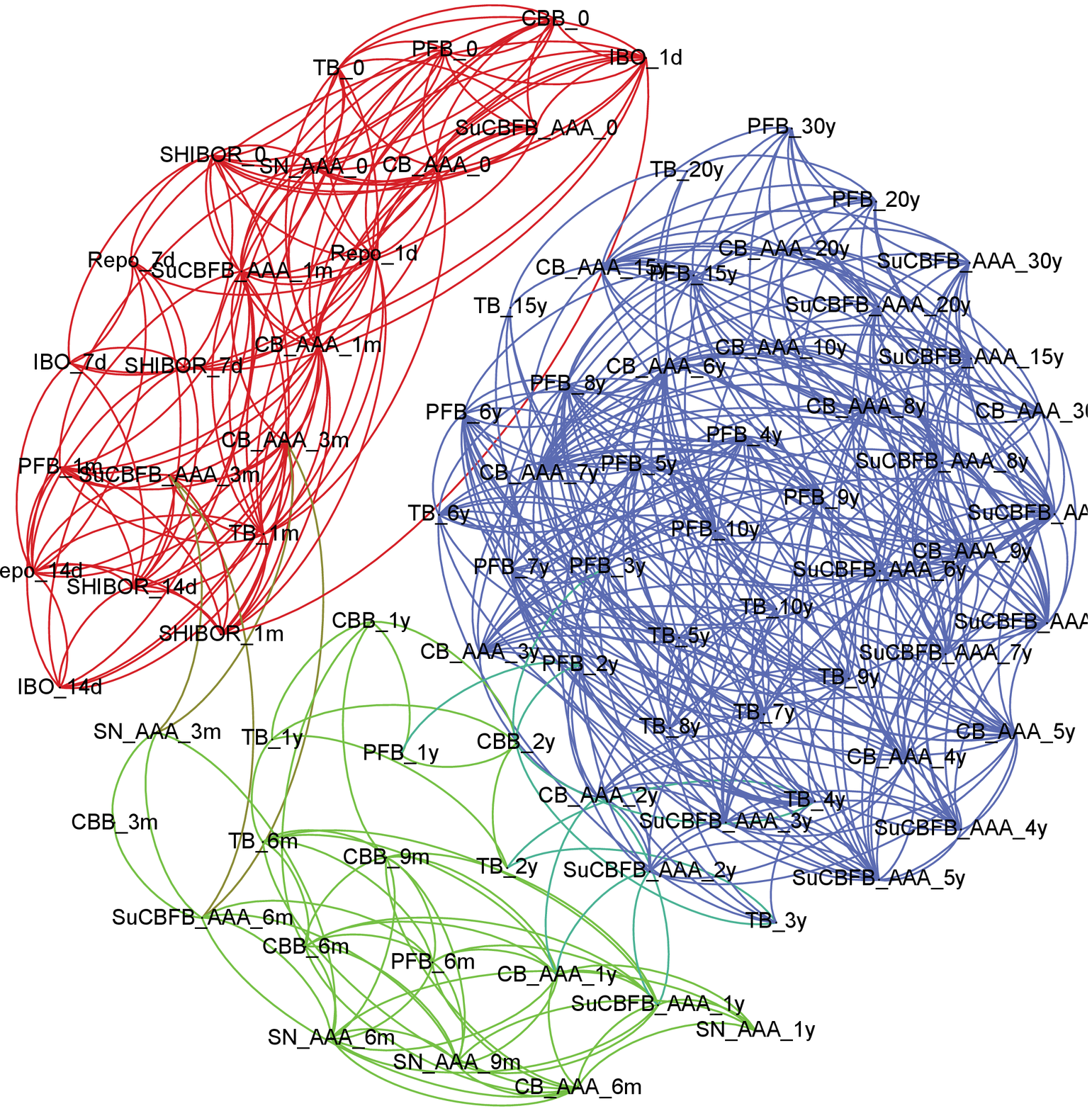}}
\hspace{0.1in}
\subfigure[$s=150$]{\includegraphics[scale=0.175]{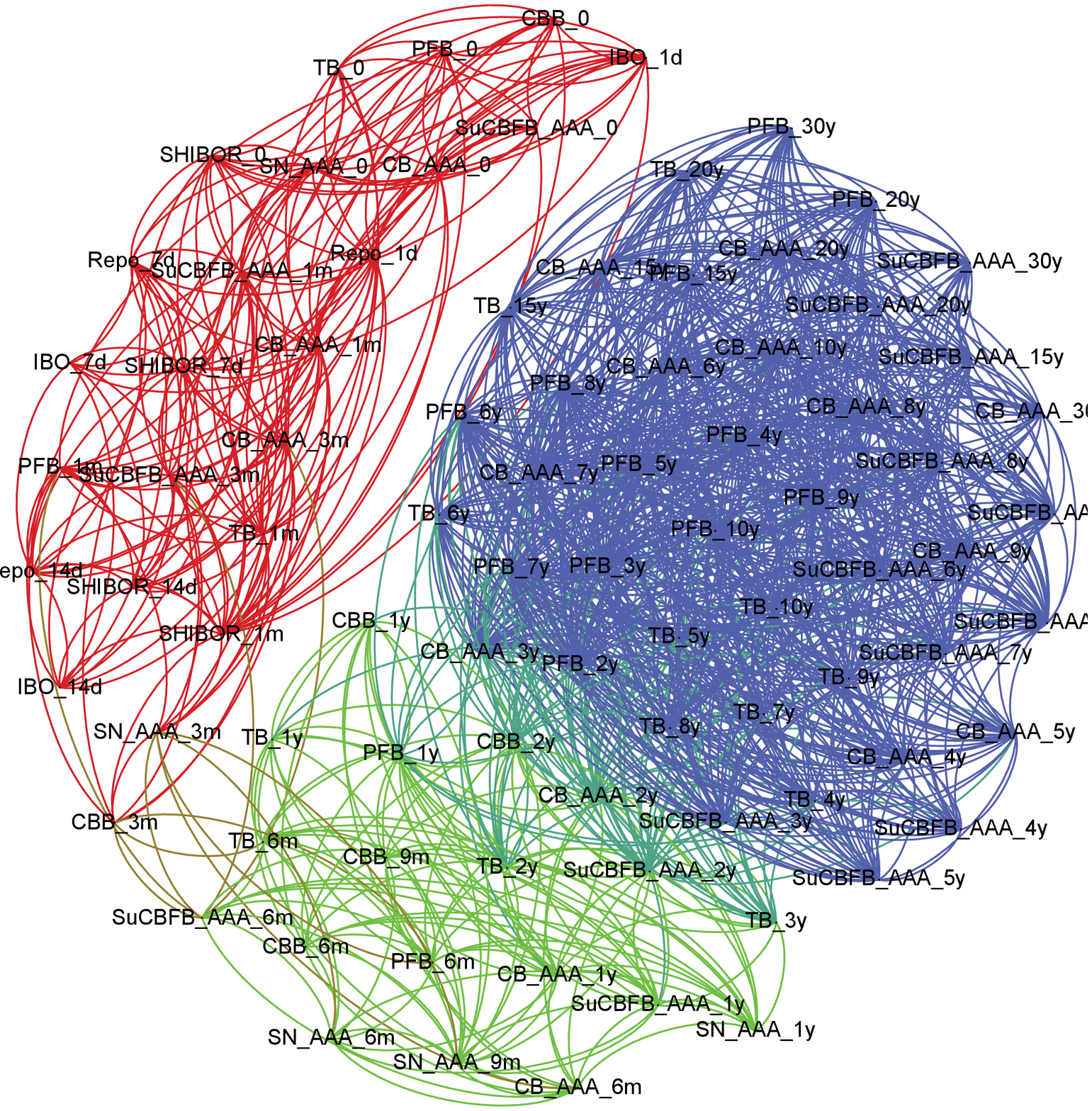}}
\hspace{0.1in}
\subfigure[$s=250$]{\includegraphics[scale=0.175]{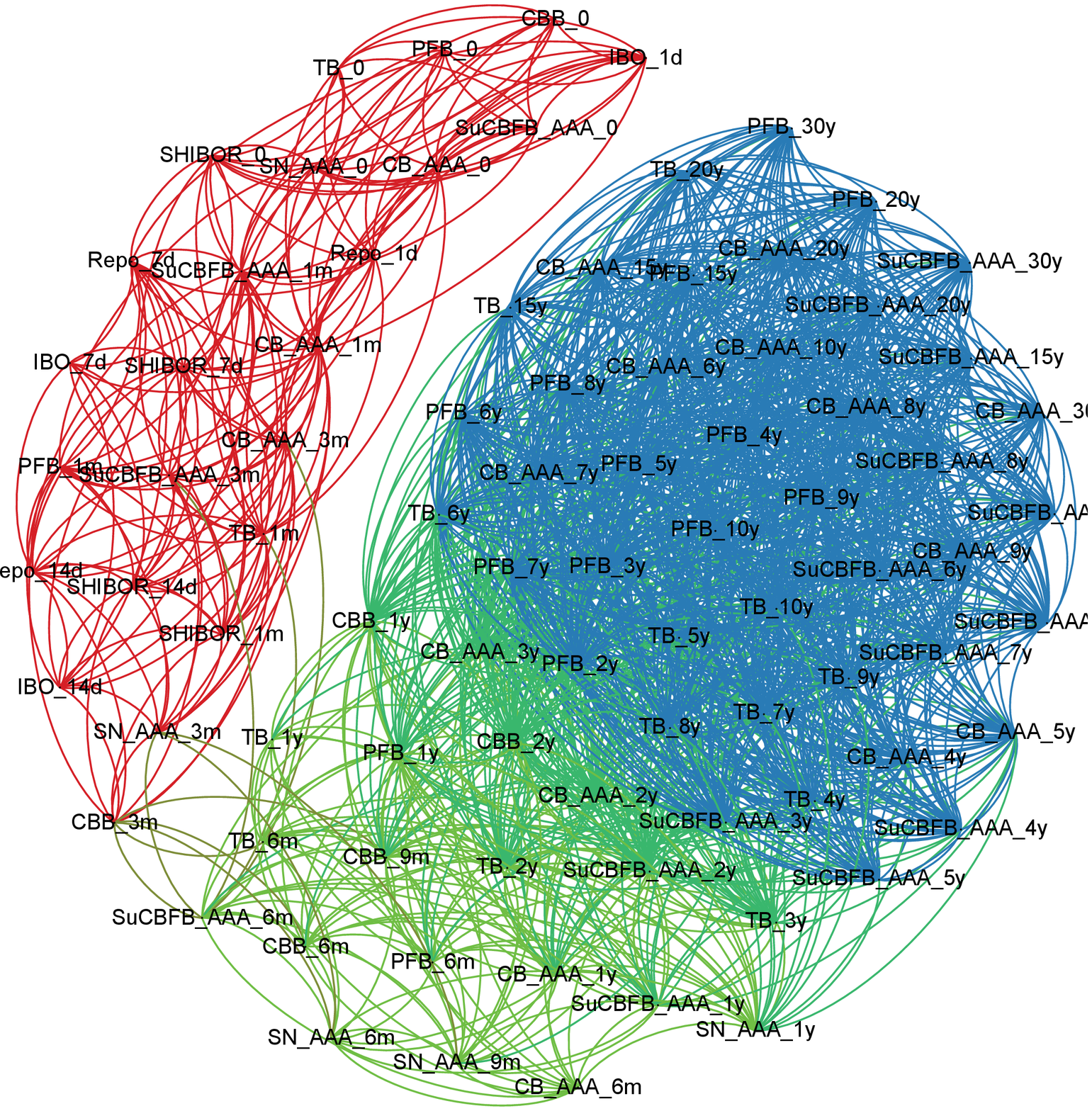}}
\caption{\footnotesize{ Long-range correlation network with windows size s=50 (left), 150 (midium), and 250 (right). An obvious segmentation structure is detected in each time scale.}}
\end{figure*}

Market segmentation theory suggests that securities including bonds can be divided into different groups according to maturities with a lack of substitution between groups for investors \cite{benson1981systematic, kidwell1983market}. Market segmentation is an important theory to explain the term structure of interest rates. Once put forward, several proofs of segmentation have been found in municipal bond market, treasury bond market and so on \cite{kidwell1983market, simon1991segmentation, hendershott1978impact}. However, to detect the segmentation structure, most literature built a regression model using the supply and demand factors of bonds, which is indirect and the supply and demand data is hard to collect. What is more, none of the literature focuses on long-range correlation between bonds. Using long-range correlation test method, our research covers this void by detecting the segmentation structure directly. Based on fractal market hypothesis (FMH), a market consists of investors with different investment horizons. In the sight of market participants, bonds with different maturities are different securities instead of substitutions. Shocks to bond market seem to diffuse within the maturity group even in the long run.\\

\subsection{ Timeline Analysis}
In the next part of this paper, we will detect the nature of long-range correlation of interest rates that changes over time. The full sample is divided into two groups according to the periods. One sub-period is from 2007 to 2012, the other is from 2011 to 2016.\\
Firstly, we detect the long-range auto-correlation via the distribution of Hurst exponents calculated by DMA process we mentioned before. The result is shown in Fig.6. Interest rates in earlier years (represented by red line) have a stronger degree of long-range dependence compared to recent years (represented by green line). The full sample period represented in blue line is shown as a benchmark in the figure. This result is in accordance with Cajueiro and Tabak \cite{Cajueiro2007Time}, who find that the degree of long-range dependence in the US interest rates has significantly decreased over time. As capital market is growing more mature, liquidity is more and more sufficient, and a shock in the market will be borne sooner.
\begin{figure}
    \centering
      \includegraphics[width=1\textwidth]{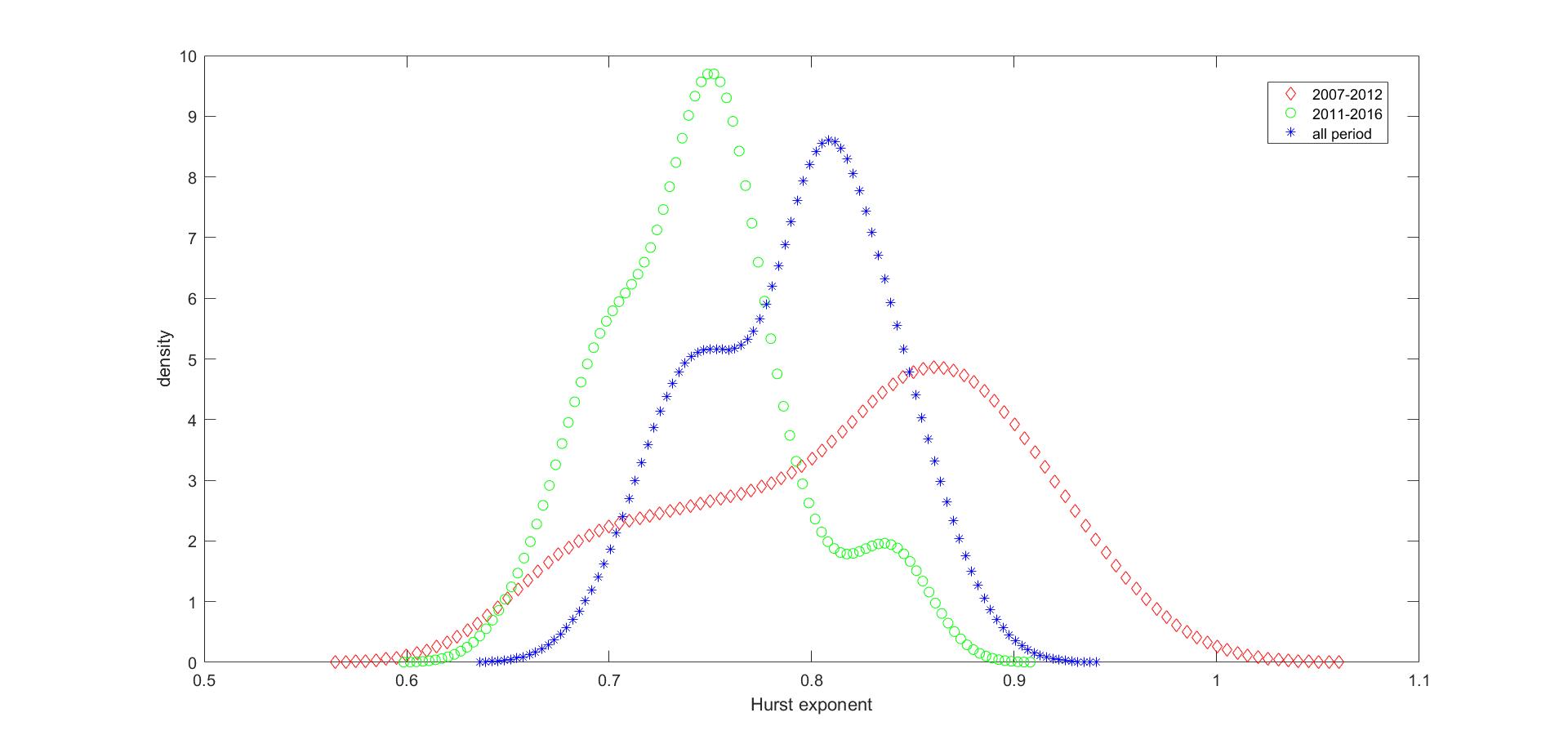}

      \caption{\footnotesize{ Distribution of Hurst exponents of earlier years (in red line) and recent years (in green line). The full sample period represented in blue line is shown in the figure as a benchmark. Interest rates have stronger long-range dependence in earlier years compared to recent years.}}

    \end{figure}

Secondly, long-range cross correlation coefficients $\rho_{DCCA} (s)$ versus time scale s are calculated for different sub-periods. After the construction of long-range correlation network, average value of weighted degree is calculated.  This can represent the level of long-range cross correlation between interest rates of the certain sub-period. The average weighted degree versus time scale s for the two sub-periods is shown in Fig.7. Represented in blue solid line, average weighted degree of network for earlier years is smaller than that of recent years (represented in orange dashed line). That is, in long run, despite weaker auto-correlation, links between interest rates have become stronger in recent years.  Asset portfolio constructing, which only takes short-range correlation into account earlier on, cannot ignore long-range correlation anymore.\\
\begin{figure}
    \centering
      \includegraphics[width=1\textwidth]{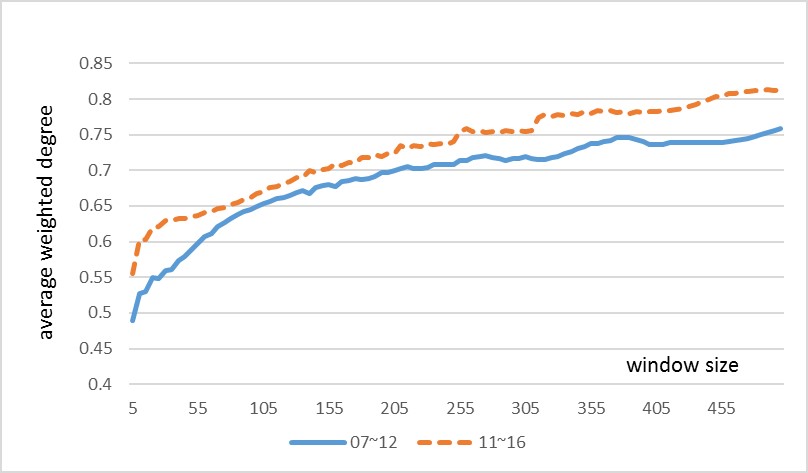}

      \caption{\footnotesize{  The average weighted degree versus time scale s for the two sub-periods. Represented in blue solid line, average weighted degree of network for earlier years is smaller than that of recent years (represented in orange dashed line). That is, in long run, despite of weaker auto-correlation, links between interest rates become stronger in recent years.}}

    \end{figure}

\section{Conclusions}
In this paper, long-range correlations between interest rates are studied in a relatively large sample of bonds. We have found a clear evidence of long-range dependence and persistence in one year scale. We have also detected the long-range correlation of some typical bonds varying over their maturities, and found some different features compared with other countries, which can be explained with the expectations of fractal market hypothesis. Then we construct a long-range cross-correlation network to study the influence structure between bonds, and find that bond market exhibits segmentation pattern even in long run. What is more, we also detect long-range auto-correlation and cross-correlation in different periods of time. The result suggests a decrease in the degree of auto-correlations but an increase in cross-correlations in recent years, which can be seen as a mark of market maturity.\\
Our contribution to current literature can be summarized as following: in theory, we combine long-range correlation method with network analysis and reveal the long-range cross-correlation structure intuitively, which covers the void of literature and provide a new way to study complex long-range correlations. Using bond market data, we also develop and provide powerful evidence to fractal market hypothesis and market segmentation theory. In practice, we see clear proof of long-range correlation in bond market, which has important implications for monetary policy purposes. Influence from a sudden shock may last for a long time. Chinese market met with a serious shortage of money in June, 2013, and overnight offered interest rate had a peak at 13.44\%. This widely volatile situation lasted for over 9 months, and interest rate was not stabilized until March, 2014.  Policy makers have to take long-range correlation into account and prepare for the long-lasting volatility risks.

\section*{Acknowledgement}

We thank for anonymous referees' suggestions and the financial support by National Science Foundation of China (Nos 71103179) and Youth Innovation Promotion Association of CAS (Grant No. 2015359) and the Open Project of Key2 Laboratory of Big Data Mining and Knowledge Management, CAS

\section*{References}

\bibliography{reference}

\end{document}